# NEW PARALLEL COMPUTING FRAMEWORK FOR RADIATION TRANSPORT CODES*†

M.A. KOSTIN[1], N.V. MOKHOV[2], K. NIITA[3]

[1]*National Superconducting Cyclotron Laboratory, 1 NSCL, East Lansing, 48824 Michigan, USA*

[2]*Fermi National Accelerator Laboratory, Batavia, 60510 Illinois, USA*

[3]*Research Organization for Information Science and Technology, Tokai-mura, Naka-gun, Ibaraki-ken 319-1106, Japan*

*Work supported by the US Department of Energy (DOE) under the grant number DE-FG02-07ER41473 and by Fermi Research Alliance, LLC under contract No. DE-AC02-07CH11359 with DOE.

†Submitted paper at *Joint International Conf. on Supercomputing in Nuclear Applications and Monte Carlo (SNA + MC2010),* October 17-21, 2010, Tokyo, Japan.

# New Parallel Computing Framework for Radiation Transport Codes

Mikhail A. KOSTIN [1*], Nikolai V. MOKHOV [2], Koji NIITA [3]

[1]National Superconducting Cyclotron Laboratory, 1 NSCL, East Lansing, 48824 Michigan, USA
[2]Fermi National Accelerator Laboratory, Batavia, 60510 Illinois, USA
[3]Research Organization for Information Science and Technology, Tokai-mura, Naka-gun, Ibaraki-ken 319-1106, Japan

A new parallel computing framework has been developed to use with general-purpose radiation transport codes. The framework was implemented as a C++ module that uses MPI for message passing. The module is significantly independent of radiation transport codes it can be used with, and is connected to the codes by means of a number of interface functions. The framework was integrated with the MARS15 code, and an effort is under way to deploy it in PHITS. Besides the parallel computing functionality, the framework offers a checkpoint facility that allows restarting calculations with a saved checkpoint file. The checkpoint facility can be used in single process calculations as well as in the parallel regime. Several checkpoint files can be merged into one thus combining results of several calculations. The framework also corrects some of the known problems with the scheduling and load balancing found in the original implementations of the parallel computing functionality in MARS15 and PHITS. The framework can be used efficiently on homogeneous systems and networks of workstations, where the interference from the other users is possible.

## I. Introduction

It is a conventional knowledge that calculations performed for shielding applications can be time consuming. A number of reasons may be responsible for that, for example the amount of shielding material in models, and comprehensive modeling of the physics processes. Various biasing and variance reduction techniques have been developed over the time to deal with this problem. Another way to mitigate the problem is using parallel computing capabilities which become easily available with rapidly growing computational resources. Most of radiation transport codes today are capable of performing calculations on parallel machines.

The parallel computing framework introduced in this paper can be used with general-purpose radiation transport codes such as MARS15[1-3] and PHITS[4]. Besides the parallel computing functionality, the framework also offers a *checkpoint facility* that can be used either in the parallel regime or in the single process regime. The framework is also proven to be more efficient than the previous implementations of parallel computing in MARS15[5] and PHITS.

*Corresponding author, E-mail: kostin@nscl.msu.edu

## II. Description of Framework

### 1. Choice of the Language and Middleware

The original idea was to make this framework as much independent of radiation transport codes it could be used with as possible. Most of these codes were developed over long periods of time, and were implemented in FORTRAN77 for the most part. FORTRAN77 data structures must be handled explicitly in a parallel code which negates the whole idea of the framework independence from any particular radiation transport code. An alternative way for the framework implementation is to segregate the parallel computing functionality into a software module written in a language that supports the mechanism of pointers. This way, any action with data structures such as packing and unpacking buffers or creating *derived types* can be done in a single loop over the pointers. C++ was chosen among several considered candidates because there are freely distributed compilers for C++, it can be mixed with FORTRAN77, and the language features allows a good code structure for the framework.

A comprehensive study of available *middleware* was performed during our previous research work[5]. To summarize, a number of candidates were considered: MPI[6,7], CORBA[8], sockets and PVM[9]. CORBA provides extensive functionality and is appropriate for distributed computing applications, but it is also relatively hard to use and its communication overhead may be significant. Sockets involve little overhead for communications but much of the necessary high-level functionality is absent. PVM has a long and successful history. The MCNP collaboration[10] however recommended MPI over PVM, having had substantial experience with both the packages. Also, given our experience, we could conclude that MPI is the best choice for this particular type of application. Besides being considered as a standard for programming parallel systems, the MPI functionality seems to match data structures and the code structure of the considered radiation transport codes quite well. It can use low level protocols such as TCP/IP which imposes little communication overhead. MPI is available for machines of all architectures – massive parallel clusters, commodity clusters and networks of workstations. Freely distributed implementations of MPI such as Open MPI[11] and MPICH[12] are available together with a number of commercial implementations with performance optimized by the vendors for their systems.

## 2. Code Architecture

The current version of the framework offers only a linear topology of *processes*. There is only one *group* of the processes with one *master* process and an arbitrary number of *worker* processes. Each process replicates the entire geometry and uses the same settings of a studied system. The parallelization is job-based, i.e. processes are running independently with different initial seeds for the random number generator. The exchange of results between the master and workers is initiated by the master according to the *scheduling* algorithm which will be described later. Besides performing the control task, the master also runs event histories. This is different from the framework originally implemented in the PHITS code. This feature is important for systems with a small number of processors.

It is quite obvious that each event in a radiation transport code is independent from other events. This differs from calculations on meshes where each new round does depend on results of previous iterations. This feature makes the processes in the framework loosely-coupled, and allows information exchange sessions (*rendezvous*) as often or rare as we choose without performance penalties. This also results in a better scalability compared to tightly-coupled calculations on meshes.

During the information exchange sessions the worker processes are inquired consecutively according to their *ranks*. In order to avoid possible interference with the running event histories, the workers probe the signals from the master in an *asynchronous mode*. A worker starts processing the next event if no signal from the master is received at the time of the probing. The master can also send out a signal to stop calculations if the required number of events has been reached. All the intermediate information is transferred from a worker to the master in a single round. The worker will pack and send a buffer with service information containing the number of processed events and the seeds for the random number generator (needed for the checkpoint facility), with the contents of arrays, and with the contents of data containers. The data containers are another part of the framework. They were developed to replace the old single-precision HBOOK histogram package[13] used in the MARS15 code. There are no limitations on the buffer size because it is dynamically allocated in the framework. All communications are performed in the MPI *standard mode* except for the signal probing mentioned above. The order of all the corresponding 'send' and 'receive' function calls is carefully matched in order to avoid a *deadlock*.

We were previously experimenting with various methods of information exchange between the processes[5]. Among the methods considered there were sending the data with MPI functions for array elements positioned contiguously in the memory, sending with the MPI functions for packing and unpacking buffers, and sending data with derived types. The data arrays in the radiation transport codes are generally positioned in a number of common blocks, therefore they are not in contiguous memory. The use of the first method would result in excessive communications, since each array should be sent separately. The other two methods can be used for data located in arbitrary places. The sending data with derived types involves some overhead to build such types, but this happens only once before the calculations are started. This is an appropriate method in case when the communications occur frequently. In our case, however, the processes are loosely coupled with information exchange sessions to be rare. In addition to that, the checkpoint facility requires functionality to pack and unpack the data to and from a buffer anyway. This is the same functionality as required for the second method. Therefore the second method is used in the framework.

## 3. Scheduling and Load Balancing

Scheduling is an important issue that is directly related to performance, scaling and fault tolerance. Since the communications are quite expensive for the current generation of commodity clusters, an obviously simple approach to the design of the framework would be reducing the amount of communications as much as possible. In the most extreme case this means collecting the information from the worker processes only once at the end of calculations. Moreover, the performance can only be good if the communication time, $T_m$ (the subscript 'm' stands for 'messages'), is much smaller than the computation time, $T_c$. On the other hand, the rendezvous must be frequent enough in order to provide some fault tolerance – it is important to be able to restart the computations from the last checkpoint if a system failure occurs. The frequent rendezvous are also useful to obtain most recent information about the calculation speed of each process in order to adjust the load of the process and to achieve a better *load balancing*.

The *scheduler* in the framework compromises between these two issues. It decides when to suspend the computation and to start a rendezvous. The decision is based on the knowledge of an estimated time needed for the rendezvous and time needed to process one event, $T_1$. The master process may wait for a response from a worker for a long time during rendezvous in case of long histories. This waiting time may significantly prolong the rendezvous. For the framework to be effective, the time between the rendezvous has to be significantly longer than $max\{T_m, T_1\}$.

In the first implementation of the parallel computing functionality in the MARS15 code[5], a worker process is terminated at a rendezvous if the number of locally processed histories combined with the number of events already collected by the master is in excess of the total number of requested events. This would most probably to happen when the jobs are close to their end. Time to the next rendezvous has to be shortened to avoid that and to use the resources more effectively. In the opposite case, the master will have to process the rest of histories by itself. This may lead to a sizeable computation time increase if the number of terminated processes is large and the balance of events is still significant. An attempt was made to take into account this effect by adjusting the two previous time conditions:

$$T = min\{100 \times max\{T_1, T_m\}, 0.8 \times T_{end}, 1\ h\}, \quad (1)$$

where T is the time to the next rendezvous, $T_{end}$ is estimated time to the end of calculations taking into account the speed of each process. The requirement of 1 h is based on a human factor. The exchange time must not exceed a sizeable fraction of a working day. This is to let people deal with potential problems in their models. The numerical factors 100 and 0.8 are parameters that can easily be changed. We have found however after several years of use that the situation when the master finishes the calculations itself still occurs. Further improvements are required despite the fact that the above scheduling algorithm offers a great load balancing.

The scheduler in the parallel computing implementation in the PHITS code does not have this problem at all. All calculations are performed in batches where each worker gets a fixed number of events to process, and the situation when the master must process an excessive number of histories is not possible. Each process 'knows' ahead of time how many events it needs to process, and stops when this number is reached. This is an example of *self-scheduling* also implemented in the MCNP code. This approach works well on homogeneous systems, but is quite inefficient on network of workstations and systems where each processor may have different performance or interference from the other users is possible.

The new framework implements features from both scheduling mechanisms. The scheduling still works as defined in (1), but each process now also knows the maximum number of events it is allowed to process. This mixed scheduling still provides a great flexibility in load balancing that allows the framework to be used on a network of arbitrary workstations, and also deals with the scheduling problems described above.

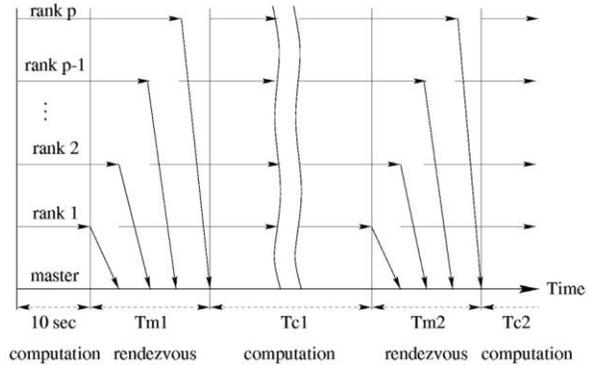

**Fig. 1** Time scheme of communications.

Figure 1 illustrates a communication time scheme. The first rendezvous occurs in a fixed time period of 10 s. The fixed time period is needed to calculate trial values for $T_1$, $T_m$ and $T_{end}$. These values are calculated at the end of each rendezvous later on, with $T_1$ calculated for each process.

### 4. Checkpoint Facility

It is not unusual that a computational job terminates prematurely and has to be restarted from the beginning. This may occur due to a variety of reasons such as power outage, problems with batch systems, etc. This may cause a significant loss of time since the computational jobs are time consuming and may take days to complete. Moreover, the radiation transport codes are under constant development. The recent code modifications sometimes make the codes unstable which may results in job terminations. Regular debugging techniques are not appropriate in this case because it may take many days to reach a problematic event.

These problems can be mitigated with a checkpoint facility that was developed as a part of the framework. The facility allows restarting calculations from a recent checkpoint and not from the very beginning. A checkpoint is represented with a file with all the information necessary to restart calculations. The information is saved into two files alternately, so that if the system fails during the framework saving the checkpoint file, then the previous file would still be available. The checkpoint facility works both in the parallel and single process regimes. The files are saved after each information exchange session in the case of parallel computation. In the single process regime, however, the user defines how often he is willing to use this functionality. Several checkpoint files can be merged into a single one, thus allowing combining results from several calculations as long as the machine word formats are the same. Currently the facility imposes no limitations in what regime a checkpoint file can be used regardless of how it was created. For example, it is possible to obtain a checkpoint file in the parallel computation regime, and to use it to restart calculations in the single process regime.

It is appropriate to say now that the existence of the checkpoint facility changes the strategy of how the calculations can be performed. The number of events needed to obtain statistically significant results is rarely known *a priori*. It is customary to estimate the required number of events using short test calculations. If these estimates are not accurate enough, the required statistical significance is not achieved, and the calculations must be restarted from the very beginning after appropriate corrections are made. The necessity to restart the calculations from the beginning may cause significant delay in obtaining results is some instances. With the checkpoint facility, however, it is not an issuer anymore, because the statistics can always be improved 'on-the-fly'.

## III. Performance Tests

Performance tests were conducted on the NSCL's DOEHPC Linux cluster. The cluster consists of 16 dual CPU nodes. Each CPU is a 2.6 GHz AMD Opteron™ 252 processor with 1024 kB of cashe memory. The nodes are equipped with 4 GB of memory and connected via a 1 Gb/s network. The MPICH2 implementation of MPI was used for the message passing. The test jobs were managed by the TORQUE Resource Manager[14].

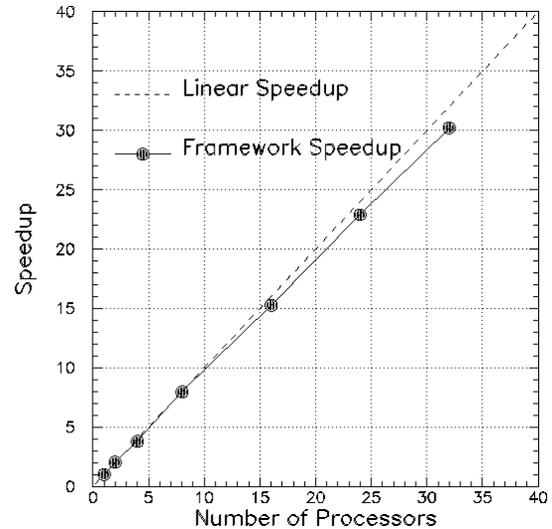

**Fig. 2**  Speedup test results.

The tests were conducted using a MARS15 model consisted of one hundred geometry zones. The performance tests measured the speedup as a function of the number of used processors. The tests demonstrated the speedup close to linear (see Figure 2). The tests could not be conducted on a larger number of processors due to the limited resources.

## IV. Framework Limitations

The framework is meant to be used in situations in which only the master process generates the output of results. In other scenarios the framework must be used with care. For instance, the user may want to generate a list of particles crossing a surface. In this case the output files with the particle lists generated by each process must be named uniquely to avoid corruption of the output data. The framework does not provide any facility in which such lists can be redirected to and saved by the master process.

The framework also assumes an unlimited input for all the nodes and processes, i.e. each process on each node must be able to read input files required by radiation transport codes. Although it is technically possible to implement a framework where only the master process reads in the input files and distribute the parameters to the worker processes, this configuration is considered outside of the scope of this work.

The authors realize that the performance tests were carried out using limited available resources. The framework speedup with currently implemented linear topology of processes may significantly degrade or even plateau out on massive parallel systems when the number of used processes is very large. This speedup degradation due to increased communication time can be significantly mitigated with considered but not yet implemented multilevel ('tree-like') topology. In the 'tree-like' topology, the master process (level 0) gathers results from a limited number of level 1 processes, which also act as masters for their own groups of processes (level 2). The number of levels is not limited.

## V. Conclusion

The new parallel computing framework was designed, implemented and integrated with the MARS15 code, and an effort is under way to integrate it with PHITS. It was tested on the NSCL small commodity cluster and demonstrated a good performance. The framework offers a good load balancing for each process so that it can be used effectively not only on homogeneous systems but also on networks of workstations. The framework performs better than the original implementations of parallel computing in MARS15 and PHITS. It also offers the checkpoint facility that can be used both in multiple and single process calculations. There is a potential to increase the efficiency of the framework through a new multilevel topology of processes, although this improvement may only be noticeable for very large parallel systems.

## Acknowledgment

The work has been supported by the US Department of Energy under the grant number DE-FG02-07ER41473.